\documentclass[12pt]{iopart}
\expandafter\let\csname equation*\endcsname\relax
\expandafter\let\csname endequation*\endcsname\relax
\usepackage{amsfonts} 
\usepackage{amsmath}
\usepackage{iopams}
\usepackage{amssymb}
\usepackage{mathtools}
\usepackage{cancel} 
\usepackage{bm}
\usepackage[usenames, dvipsnames]{color} 
\usepackage{dcolumn}
\usepackage{epic} 
\usepackage{epsfig}
\usepackage{feynmp}
\usepackage{graphicx}
\usepackage{grffile}
\usepackage{comment}
\usepackage{float}
\usepackage[breaklinks,colorlinks = true,linkcolor = red,urlcolor=blue,citecolor=red]{hyperref}
\usepackage{mathrsfs}
\usepackage{soul}
\usepackage{subfigure}
\usepackage{wrapfig}
\usepackage{xy} 
\usepackage{xcolor}
\usepackage{amsmath,amssymb,amsfonts,amsthm}
\usepackage[ascii]{inputenc}
\usepackage{makecell}
\usepackage{cite}
\allowdisplaybreaks
\usepackage{etoolbox}
\usepackage[normalem]{ulem}

\newcommand*\mathinhead[2]{\texorpdfstring{${#1}$}{#2}}

\makeatletter
\newrobustcmd{\fixappendix}{%
  \patchcmd{\l@section}{1.5em}{7em}{}{}%
  \patchcmd{\l@subsection}{2.3em}{7em}{}{}%
}
\makeatother

\graphicspath{{images/}{../Figures}}

\begin{document}

\title{Impact of dephasing probes on incommensurate lattices}
\author{Bishal Ghosh}
\address{Department of Physics,
		Indian Institute of Science Education and Research, Pune 411008, India}
		\vspace{0.2cm}
\address{Department of Physics, The University of Texas at Dallas, Richardson, Texas 75080, USA}
  \author{Sandipan Mohanta}
  \address{Department of Physics,
		Indian Institute of Science Education and Research, Pune 411008, India}
\author{Manas Kulkarni}
\address{International Centre for Theoretical Sciences, Tata Institute of Fundamental Research,
Bangalore 560089, India}
\author{Bijay Kumar Agarwalla}
\address{Department of Physics,
		Indian Institute of Science Education and Research, Pune 411008, India}

\begin{abstract}
We investigate open quantum dynamics for a one-dimensional incommensurate Aubry-Andr\'{e}-Harper lattice chain, a part of which is initially filled with electrons and is further connected to dephasing probes at the filled lattice sites. This setup is akin to a step-initial configuration where the non-zero part of the step is subjected to dephasing. We investigate the quantum dynamics of local electron density, the scaling of the density front as a function of time both inside and outside of the initial step, and the growth of the total number of electrons outside the step. We analyze these quantities in all three regimes, namely, the de-localized, critical, and localized phases of the underlying lattice. Outside the initial step, we observe that the density front spreads according to the underlying nature of single-particle states of the lattice, for both the de-localized and critical phases.  For the localized phase, the spread of the density front hints at a logarithmic behaviour in time that has no parallel in the isolated case (\emph{i.e.}, in the absence of probes).  Inside the initial step, due to the presence of the probes, the density front spreads in a diffusive manner for all the phases. This combination of rich and different dynamical behaviour, outside and inside the initial step, results in the emergence of mixed dynamical phases. While the total occupation of electrons remains conserved, the value outside or inside the initial step turns out to have a rich dynamical behaviour. Our work is widely adaptable and has interesting consequences when disordered/quasi-disordered systems are subjected to a thermodynamically large number of probes. 
\end{abstract}
\date{\today}
\maketitle
\section{Introduction}
\label{intro} 
Understanding and manipulating the quantum dynamics of open systems is of significant interest both from a fundamental and a practical perspective \cite{breuer2002theory,carmichael2009statistical, weiss2012quantum,RevModPhys.88.021002,davies1976quantum,agarwal2012quantum, rotter2015review, RevModPhys.93.015008, 10.21468/SciPostPhysLectNotes.68}. One of the standard ways to investigate open quantum systems is via the method of quantum master equations, such as the Lindblad equation \cite{lindblad1976generators,manzano2020short,PhysRevA.93.062114,tupkary2022fundamental,PhysRevA.107.062216,BauerNotes,maniscalco2004lindblad, trivedi2022filling}. This involves tracing out the degrees of freedom of the environment under certain well-known approximations, resulting in an effective prescription for the dynamics of the reduced density matrix of the system \cite{carmichael2009statistical,breuer2002theory}. In the Lindblad framework (and more generally in open quantum systems), the impact of the environment often is of two types: (i) dissipator-like \cite{dissipator_2,Segal_3,KOROL2018396,Dissipator_agarwalla, dissipator_3,dissipator_4,dissipator_5,dissipator_6,krapivsky2019free, krapivsky2020free} and (ii) dephasing-like \cite{dephasing_num,dephasing_3,Dephasing_landi, Sels_non_gaussian, dephasing_4,dephasing_5,PhysRevB.93.094205, kulkarni2012towards, kulkarni2013full}. The former encodes potential exchange of particles between the system and the environment whereas the latter encodes energy/heat exchange but conserving the total number of particles in the system. Our work revolves around the role of dephasing probes. 

The impact of dephasing probes on the quantum dynamics of a lattice that is filled with particles has been an active area of research \cite{dephasing_num,dephasing_3,Dephasing_landi, Sels_non_gaussian, dephasing_4,dephasing_5}. It is also important to note that often such dephasing probe scenario is studied by adding a stochastic noise coupled to a number-conserving term in the Hamiltonian of the lattice \cite{Sels_non_gaussian, dephasing_4, dephasing_5, SciPostPhys1,singh2023noise,knap_2}. For example, it was recently shown that for a one-dimensional disordered lattice, displaying Anderson localization, when subjected to a local dephasing, density excitation spreads logarithmically in time \cite{SciPostPhys1}. Very recently, this study involving dephasing probes was extended for the incommensurate Aubrey-Andre-Harper (AAH) lattice model \cite{singh2023noise}. In particular, the authors of  Ref.~\cite{singh2023noise} explore the role of single and multiple dephasing probes on incommensurate lattices. Note that, having a thermodynamically large number of probes leads to a diffusive behavior irrespective of the underlying isolated AAH system \cite{aubry1980analyticity, PGHarper_1955, PhysRevE.96.032130,PhysRevB.100.085105,PhysRevLett.110.180403,Archak_AAH} being in a de-localized, critical, or in localized phase. This naturally leads to the intriguing prospect of mixed dynamical phases in such an engineered open system, which is one of the central themes of our work. The possible coexistence of such mixed phases is far from obvious and relies on a non-trivial geometry and arrangement of a thermodynamically large number of probes. 

\begin{figure}[h]
\centering
\includegraphics[width=1.0\textwidth]{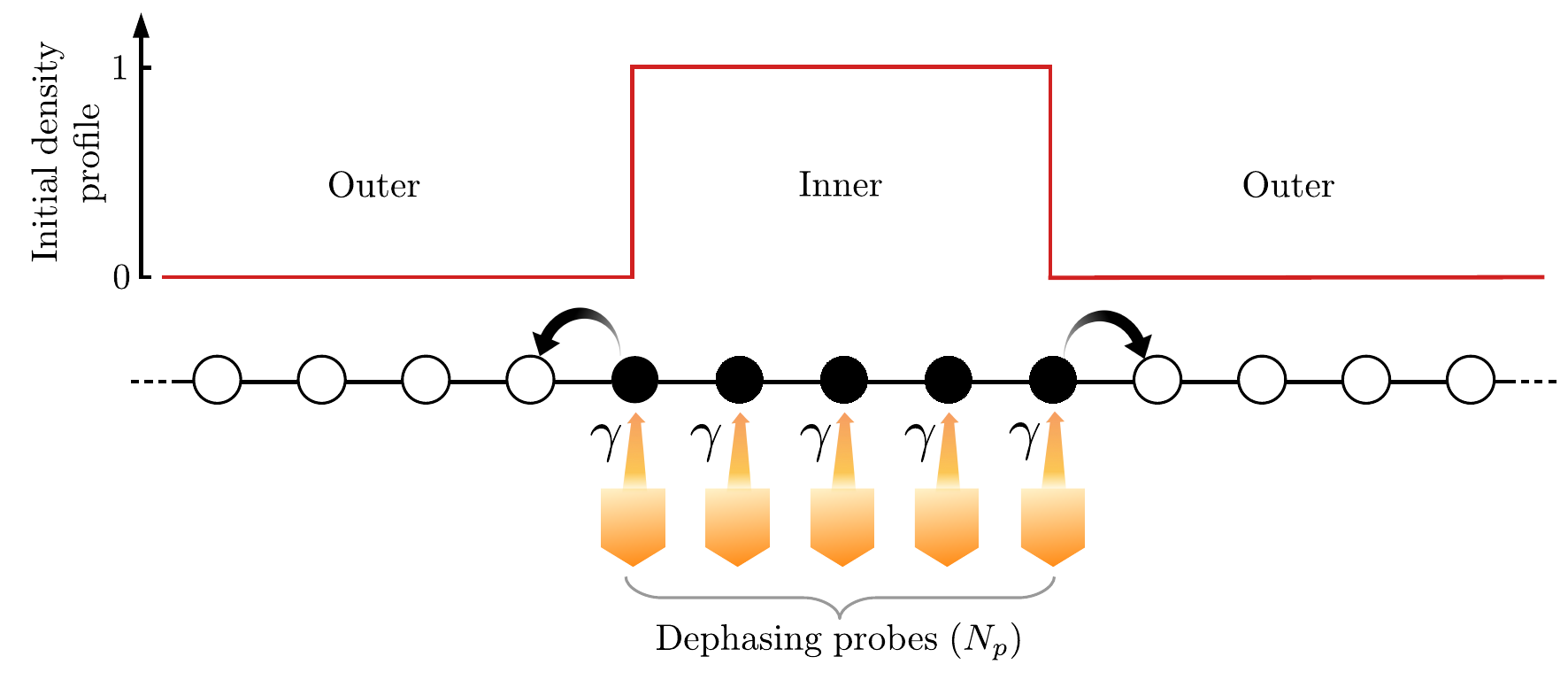}
\caption{ Schematic of a one-dimensional fermionic lattice where the central region of the lattice is initially fully filled with electrons and is represented here by filled black circles. The other lattice sites are initially empty and represented by hollow circles. This results in a step-like initial condition for the spatial electron density. This is schematically represented by a red line. This initial step function demarcates the regimes as outer-inner-outer. The arrows between the filled and the empty regions indicate the potential hopping that can give rise to interesting quantum dynamics.  At each filled lattice site (\emph{i.e.}, non-zero part of the step function), a dephasing probe is attached with homogeneous coupling strength $\gamma$. The total number of dephasing probes in the setup is $N_p$.} 
\label{Schematic}
\end{figure}

In this work, we consider a one-dimensional AAH chain, that offers various kinds of single-particle states and thereby different phases. The central region of the lattice chain is initially fully filled by electrons, resulting in a density profile reminiscent of a step-like initial configuration.  We investigate the impact of dephasing probes, that are connected at all the initially filled  (\emph{i.e.}, non-zero part of the step configuration) sites. A schematic of our setup is provided in Fig.~\ref{Schematic}. We focus on the quantum dynamics of (i) the local electron density profile, (ii) the scaling of the density front as a function of time both inside and outside of the initial step, and (iii) the growth of the total number of electrons outside the step-like configuration. We analyze these quantities in all three regimes of the underlying AAH lattice, namely, the de-localized, critical, and localized phases. Below we summarize some of our central findings:
\begin{itemize}
\item The setup in schematic Fig.~\ref{Schematic} offers a unique way to realize mixed (inner-outer) dynamical phases, \emph{i.e.}, diffusive-ballistic (Fig.~\ref{de-loc-fig}), diffusive-diffusive (Fig.~\ref{cric-fig}), or diffusive-logarithmic (Fig.~\ref{localized-plot}) when the underlying AAH lattice is in de-localized, critical, or localized regime, respectively. The local spatial density profile serves as a remarkable observable for exploring these mixed dynamical phases. 

\item Although the total occupation of electrons within the lattice is a conserved quantity, even when dephasing probes are present, the number of electrons in the inner and the outer regimes are not independently constant and exhibits intriguing dynamics. In particular, the growth of occupation within the outer regime, denoted by $N_{\rm outer}(t)$, displays markedly distinct behavior across the delocalized, critical, and localized phases. This is shown in Fig.~\ref{de-loc-fig-outer}, Fig.~\ref{cric-fig-outer}, and Fig.~\ref{loc-fig-outer}.

\item Furthermore we explicitly demonstrate the interesting differences in the time dynamics of $N_{\rm outer}(t)$ in the presence of a thermodynamically large number of dephasing probes as opposed to the absence of the probes.  This is shown in Fig.~\ref{de-loc-fig-outer} (delocalized), Fig.~\ref{cric-fig-outer} (critical), and Fig.~\ref{loc-fig-outer} (localized).

\end{itemize}

We organize the paper as follows: In Sec.~\ref{S1}, we first discuss the model and provide the details of the method used to study the observables of interest. In Sec.~\ref{S2}, we provide the results for (i) local spatial density profile, (ii) the spread of the density front, and (iii) the fate of quantum dynamics for the total number of electrons in the outer regime $N_{\rm outer}(t)$. Conclusion along with an outlook is presented in Sec.~\ref{sum}. Certain details are provided in the appendix. 

\section{Setup and Method}
\label{S1}
In this work, we focus on understanding the quantum dynamics of a setup (see schematic Fig.~\ref{Schematic}) where the central region of the  Aubry-Andr\'{e}-Harper (AAH) lattice is initially fully filled and is subjected to dephasing probes.
We write down the Hamiltonian for the AAH lattice as,
\begin{equation}
    {H}_S =\! \sum_{i = 1}^{N}\varepsilon_i \, c_i^{\dagger}\, c_i +  J  \sum_{i = 1}^{N-1} \, \big( c_{i+1}^{\dagger}\, c_i + c_i^{\dagger} \, c_{i+1}\big),
\end{equation}
where $N$ is the total number of lattice sites, $c_i^{\dagger}$ and $c_i$ are the creation and the annihilation operators for an electron at the $i$-th lattice site. $J$ represents the nearest-neighbor hopping strength and $\varepsilon_i$ represents the onsite potential strength  which for the AAH lattice is given as \cite{aubry1980analyticity, PGHarper_1955},
\begin{equation} \label{eq:2} 
\varepsilon_i = 2\,\lambda \, \cos[2 \pi  b \, i + \phi].
\end{equation}
Here  $\lambda$ is the strength of the quasiperiodic lattice potential, $b$ is an irrational number, and $\phi$ is the phase factor that lies between $[0,2\pi]$. For our work, we consider the value of $b$ as the golden mean of the Fibonacci sequence, \emph{i.e.}, $b\!=\!(1 + \sqrt{2})/5$. Interestingly, depending on the ratio of the quasiperiodic potential strength $\lambda$ to the hopping parameter $J$,  \emph{i.e.}, $\lambda/J$, the nature of the single-particle eigenstates for the AAH model changes drastically\cite{aubry1980analyticity, PGHarper_1955}. For $\lambda /J \!<\! 1$, all the single-particle eigenstates are delocalized whereas for $\lambda /J \!>\! 1$ all the single-particle states are exponentially localized, and $\lambda /J \!=\! 1$ represents the critical point for delocalized to localized transition. At this critical point, the single-particle states are neither localized nor delocalized. 

We next discuss the implementation of the dephasing probes that are attached at the initially filled lattice sites. In order to mimic the impact of these probes on the lattice, we model the dynamics for the system (lattice chain) via a Lindblad quantum master equation. We write the governing equation for the reduced density operator $\rho_S$ of the lattice as,
\begin{equation}
\dot{{\rho}}_S= -i \big[H_S, \rho_S\big]
+ \sum_{i = 1}^{N_p}\gamma_i \big( L_{i} \rho_S L_{i}^{\dagger} - \frac{1}{2} \big\{ L_{i}^{\dagger} L_{i}, \rho_S \big\}  \big).
\label{QME}
\end{equation}
The first term here represents the unitary evolution due to the lattice. The second term emulates the effect of the interaction with the dephasing probes. $L_i$ represents the jump operator for the $i$-th lattice site. For the dephasing probe considered here,  we set
\begin{equation}
L_{i} = {c}_i^{\dagger} {c}_{i} \equiv \hat{n}_{i}\,,
\end{equation}
where $n_i$ is the electron density at the $i$-th lattice site. Such an interaction preserves the total number of electrons in the lattice.  $\gamma_i$ in Eq.~\eqref{QME} represents the coupling strength between the system and the probe. $N_p$ refers to the total number of dephasing probes attached to the lattice.

Our initial setup is as follows: The central region of the AAH lattice is initially (at $t\!=\!0$) fully filled with electrons. Dephasing probes are connected at all the filled lattice sites. Such an initial condition is akin to a step-initial configuration and divides the lattice into left and right outer regimes (devoid of electrons) and an inner regime (filled with electrons). This is clearly shown in schematic Fig.~\ref{Schematic}. In order to investigate the quantum dynamics of this setup, we write down the equation of motion for the two-point correlation matrix following the Lindblad QME [Eq.~\eqref{QME}] as, 
\begin{eqnarray}
        \frac{d}{dt} \langle{c}_{i}^{\dagger}{c}_{j}\rangle &=& i\big(\langle{c}_{i+1}^{\dagger}{c}_{j}\rangle \!+ \langle{c}_{i\!-\!1}^{\dagger}{c}_{j}\rangle \!-\! \langle{c}_{i}^{\dagger}{c}_{j\!-\!1}\rangle \!-\! \langle{c}_{i}^{\dagger}{c}_{j+1}\rangle\big)  \nonumber  \\
        &-&\!\frac{\gamma}{2} \sum_{l = 1}^{N_p} \, \big(\delta_{il}\langle{c}_{i}^{\dagger}{c}_{j}\rangle + \delta_{jl}\langle{c}_{i}^{\dagger}{c}_{j}\rangle\big) +   \gamma \sum_{l = 1}^{N_p} \, \delta_{li}\,\delta_{lj}\langle{c}_{l}^{\dagger}{c}_{l}\rangle.
        \label{corr}
\end{eqnarray}
It is important to note that while the QME in Eq.~\eqref{QME} involves local quartic-type dephasing interaction $L_i \!=\! \hat{n}_i$, this being Hermitian in nature, facilitates the closure of the equation of motion for a two-point correlation function without requiring any higher order terms \cite{Sels_non_gaussian}, as given in Eq.~\eqref{corr}.  
We introduce the following definitions,  
\begin{eqnarray}\label{eq:coherence}
C_{ij} =\langle{c}_{i}^{\dagger}{c}_{j}\rangle, \quad 
D_{ij} = \frac{\gamma}{2}\sum_{l = 0}^{N_p}\delta_{il}\delta_{jl}, \quad 
P_{ij} = \gamma\sum_{l = 0}^{N_p}\delta_{il}\delta_{jl}\langle{c}_{l}^{\dagger}{c}_{l}\rangle.
\end{eqnarray}
As a consequence, we can rewrite Eq.~\eqref{corr} in a matrix form as,
\begin{equation}
    \frac{d}{dt} {C} = -i \big[H_S, C \big]  - \big \{C, D \big\} + P,
    \label{eq-matrix-form}
\end{equation}
where, note that all the matrices appearing in Eq.~\eqref{eq-matrix-form} are of size $N \times N$, with $N$ being the total number lattice sites. $D$ and $P$ are the diagonal matrices, defined in Eq.~\eqref{eq:coherence}, and capture the effect of the dephasing probes.
Alternatively, one can write Eq.~\eqref{eq-matrix-form}  as 
\begin{equation}
    \frac{d}{dt} C = -i H_{\rm e}  C + i  C  H_{\rm e}^\dagger + P,
    \label{central}
\end{equation}
where we define the effective non-Hermitian Hamiltonian 
\begin{equation}
H_{e} = H_S - i D.
\end{equation} 
Eq.~\eqref{central} is the central equation that we utilize to study the quantum dynamics for local density, defined as
\begin{equation}
n_i(t)= \langle c_i^{\dagger} c_i \rangle, 
\label{local-density}
\end{equation}
and the growth of the total number of electrons outside the initial step-like configuration (say the right outer), 
\begin{equation}
N_\mathrm{ outer}(t)= \!\sum_{i=N_p+1}^{N} \langle c_i^{\dagger} c_i \rangle.\label{outer}
\end{equation}
We use the Runge-Kutta 4th-order integrator to numerically solve Eq.~\eqref{central} and subsequently extract the quantities of interest, given in Eqs.~\eqref{local-density} and \eqref{outer}. In the next section, we present our findings. 

\section{Results for the Aubry-Andr\'{e}-Harper (AAH) model}
\label{S2}
Before discussing our numerical findings, we recall a few salient features of the underlying isolated AAH lattice. This lattice hosts three distinct regimes, namely de-localized ($\lambda/J \!<\!1$), critical ($\lambda/J\!=\!1$), and localized ($\lambda/J \!>\!1$). Several open quantum versions of this setup have also been a subject of intense research interest \cite{PhysRevE.96.032130, Archak_AAH,singh2023noise,CUI2022127778}. {There has also been enormous interest in non-Hermitian versions of the AAH lattice \cite{longhi2019topological, zeng2017anderson, longhi2021phase, longhi2019metal,zeng2020topological}.}
However, the role of dephasing probes in generating mixed phases has not been addressed. This is now investigated in subsections \ref{sub-delo} (de-localized), \ref{sub-cric} (critical), and \ref{localized} (localized). In particular, we focus on two quantities of interest, namely, the local density profile $n_i(t)$, as defined in Eq.~\eqref{local-density}, and the total number of electrons in the outer regime $N_{\rm outer}(t)$, as defined in Eq.~\eqref{outer}. It is to be noted that while we focus on $N_{\rm outer}(t)$ in the outer regime, it is the same as the number of depleted electrons in the inner regime, as the quantum dynamics progresses.  For all our numerical simulations we choose $\gamma\!=\!J\!=\!1$, without any loss of generality. We average over $100$ phase ($\phi$) realizations of the AAH lattice.
\begin{figure}[t]
\centering
\includegraphics[scale=0.6]{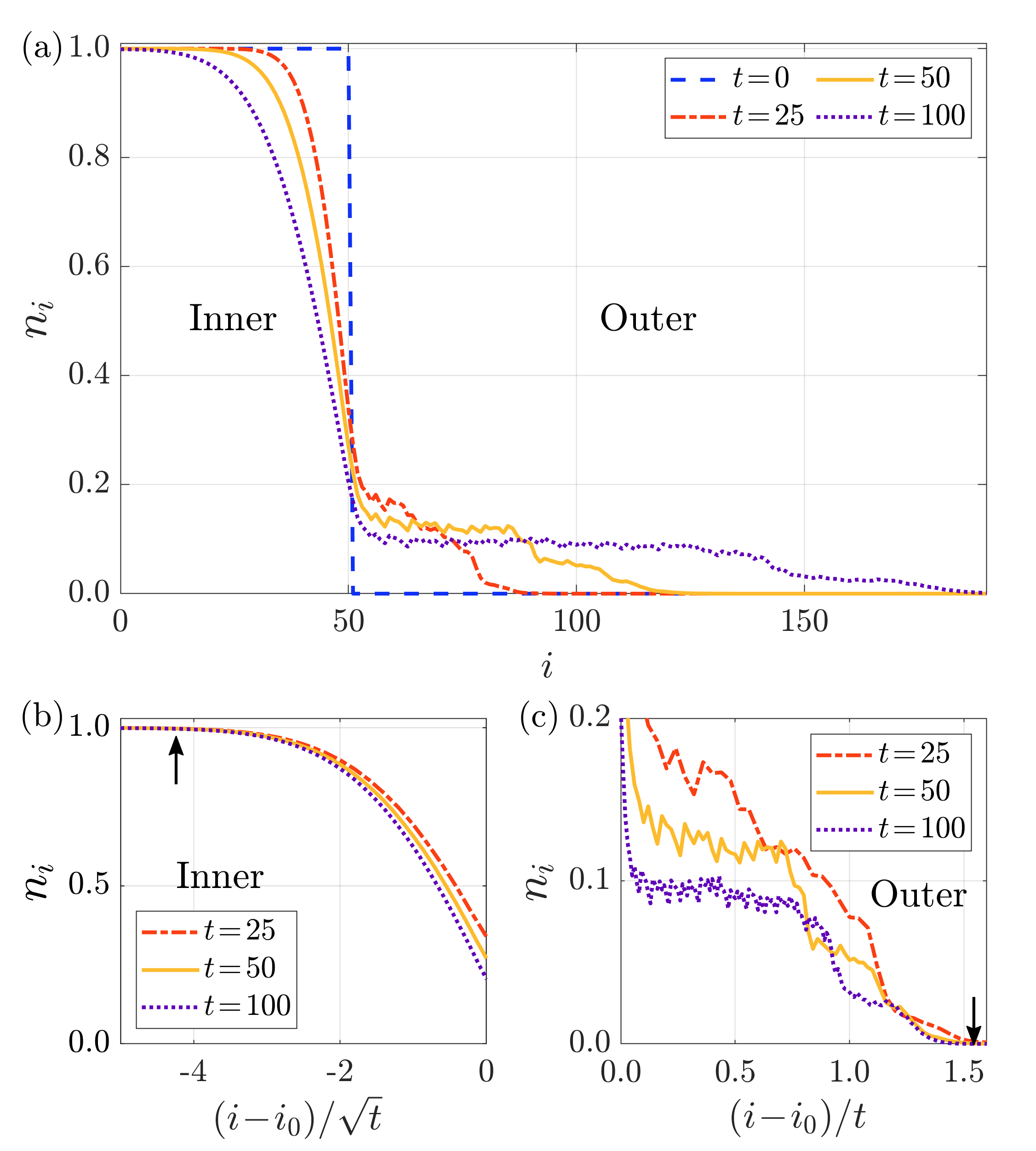}
\caption{$\bm{[}$\textbf{Delocalized regime}$\bm{]}$ (a) Plot for local density profile $n_i(t)$, given in Eq.~\eqref{local-density}, as a function of lattice site $i$ for different time snapshots for the AAH lattice in delocalized regime, $\lambda/J \!<\!1$, in the presence of dephasing probes.  (b) Density profile in the inner regime (\emph{i.e.}, sites that are coupled to dephasing probes) with the $x$-axis scaled as $(i\!-\!i_0)/\sqrt{t}$ where $i_0$ is the location of the right edge of the step. The perfect collapse of the front (indicated by the upward arrow) indicates the diffusive spreading of the depletion front in the inner regime. (c) Density profile in the right outer regime (\emph{i.e.}, sites that are not coupled to dephasing probes) with the $x$-axis scaled as $(i\!-\!i_0)/{t}$. The perfect collapse of the front (indicated by the downward arrow) indicates the ballistic spreading of the growing front in the outer regime. For all the plots, we choose the lattice size $N\!=\!501$ with $101$ filled lattice sites which are subjected to $N_p\!=\!101$ dephasing probes. We take $\lambda\!=\! 0.5$ and $J$ is always taken as $1$. The blue dashed line represents the initial density profile for the lattice. Unless otherwise mentioned, in this paper, due to the left-right symmetric nature (recall schematic Fig.~\ref{Schematic}), we display results for the right half of the setup. }
\label{de-loc-fig}
\end{figure}

\begin{figure}
\centering
\includegraphics[width=1.0\textwidth]{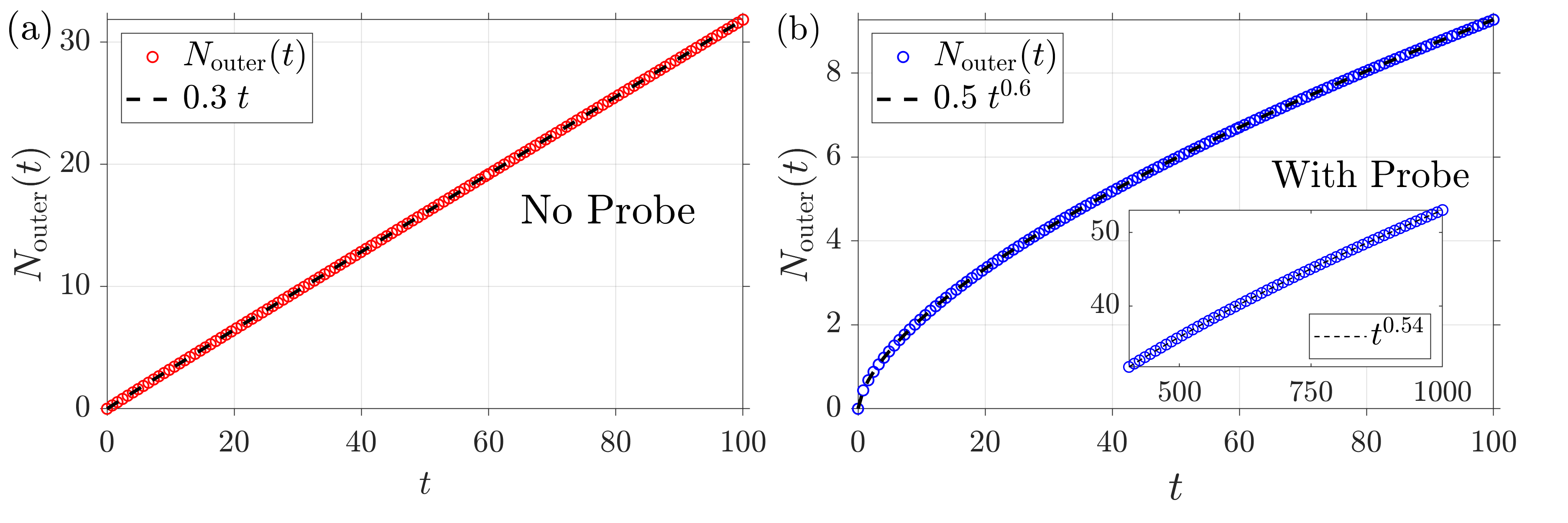}
\caption{$\bm{[}$\textbf{Delocalized regime}$\bm{]}$\, (a) Plot for the total number of electrons $N_{\rm outer}(t)$, given in Eq.~\eqref{outer}, as a function of $t$ in the outer regime in the absence of the dephasing probes for $\lambda\!=\!0.5$. It is clearly seen that $N_{\rm outer}(t)$ has a power-law behavior in time, \emph{i.e.}, $N_{\rm outer}(t) \!\propto\! t^b$ with $b\! \sim\! 1$ for no probe case. (b) Plot for the total number of electrons $N_{\rm outer}(t)$, given in Eq.~\eqref{outer}, as a function of $t$ in the outer regime in the presence of the dephasing probes for $\lambda\!=\!0.5$. In this case, we find the exponent $b\! \,\approx \!\,0.6$  when a thermodynamically large number of probes are present. The inset in the right figure demonstrates close to $\sqrt{t}$ behaviour for $N_{\rm outer}(t)$ for large $t$.}
\label{de-loc-fig-outer}
\end{figure}
  
\subsection{Delocalized phase \mathinhead{\lambda/J <1}{}:}
\label{sub-delo}
In Fig.~(\ref{de-loc-fig}) we present our findings of the quantum dynamics in the case where the isolated system is in the delocalized phase, \emph{i.e.},  
$\lambda/J \!<\!1$. From schematic Fig.~\ref{Schematic}, recall that our initial setup is a 
lattice chain that is fully filled with electrons at the central region of the lattice and therefore resembles a step-initial configuration. Furthermore, all the initial filled sites are impacted by a thermodynamically large number of dephasing probes. In other words, each filled lattice site is attached to a dephasing probe. In Fig.~(\ref{de-loc-fig})(a) we plot the local density profile $n_i(t)$, given in Eq.~\eqref{local-density}, as a function of lattice site $i$ for different time snapshots. It is clear from this figure that the system hosts two distinct dynamical phases, represented by the inner and outer regimes. The inner regime, where the dephasing probes are attached, has a propagating depletion front. Likewise, the outer regime, \emph{i.e.}, the regime devoid of probes, facilitates a propagating growing front. 
Interestingly, at any given time snapshot, a plateau-like structure in the outer regime appears. As time progresses the height of this plateau keeps decreasing and the length of the plateau keeps increasing until the finite size effects due to boundaries kick in. In the steady state, due to the finite size $N$ of the lattice, one would anticipate a uniform density distribution across it.
In other words, the reduced density matrix $\rho_s \!\propto\! I$ is a steady-state solution for Eq.~\eqref{QME}.
The value of the uniform steady-state density is given by the ratio between the total number of electrons present in the lattice and the total number of sites, which for this case is $N_p/N \approx 0.2$. It is also to be noticed that the outer regime has an oscillatory profile which is rooted in the intricate interplay between the population $\langle c_i^{\dagger} c_i \rangle $ and the coherence $\langle c_i^{\dagger} c_j \rangle$, with $i \!\neq\! j$, as can be seen in Eq.~\eqref{corr}. On the other hand, in the inner regime, the dynamics is entirely governed by the population and the coherences are destroyed by the dephasing probes. Hence the inner regime is devoid of oscillations (see \ref{appendix_A} for the details regarding coherences or lack thereof).

\begin{figure}[t]
\centering 
\includegraphics[scale=0.6]{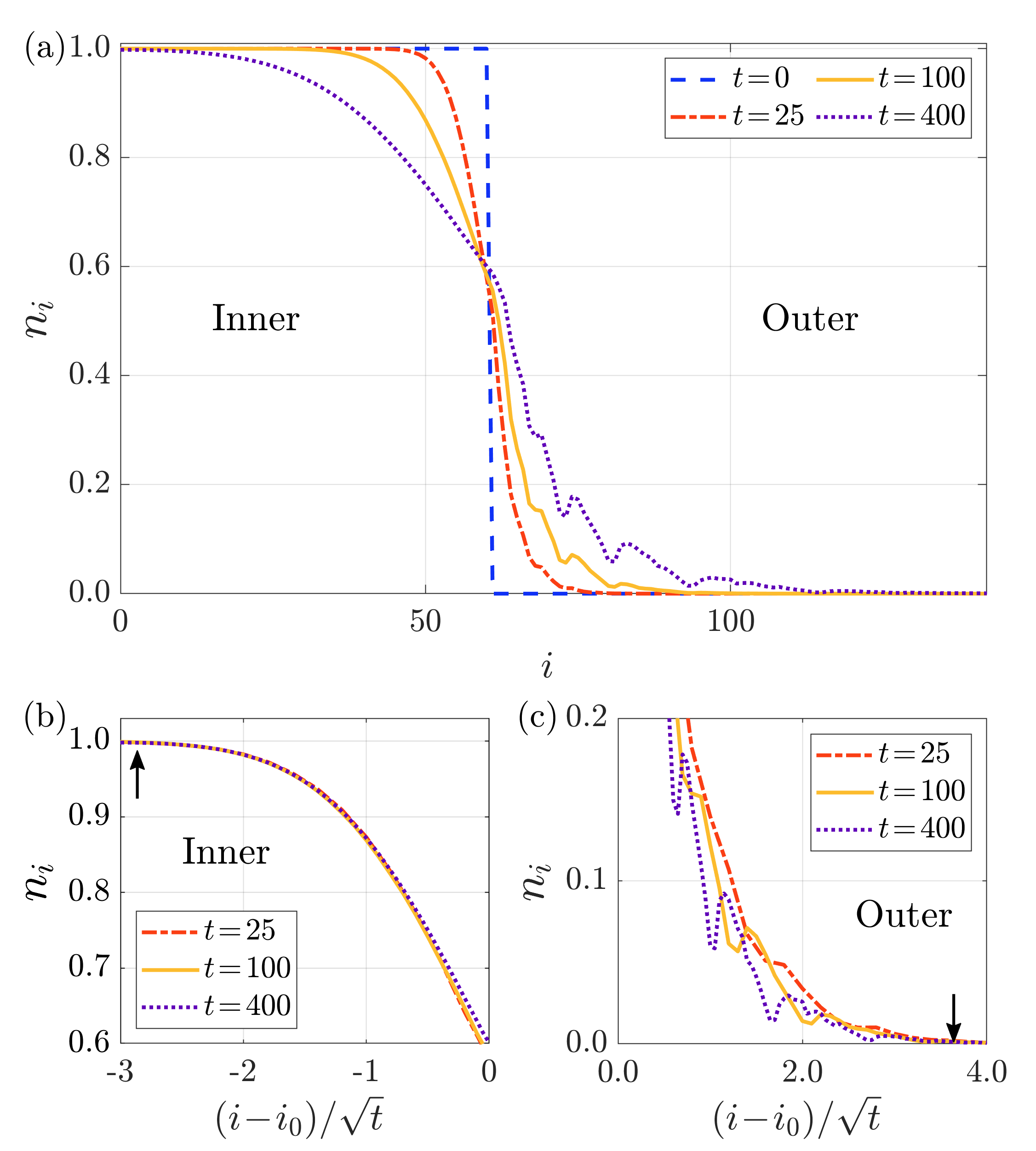}
\caption{$\bm{[}$\textbf{Critical regime}$\bm{]}$ (a) Plot for the local density profile $n_i(t)$, given in Eq.~\eqref{local-density}, as a function of lattice site $i$ for different time snapshots for AAH lattice in the critical phase $\lambda/J\!=\!1$ in the presence of dephasing probes. (b) Density profile in the inner regime (\emph{i.e.}, sites that are connected to dephasing probes) with the $x$-axis scaled by $(i\!-\!i_0)/\sqrt{t}$ where $i_0$ is the location of the right edge of the step. The perfect collapse of the density front (indicated by the upper arrow) indicates diffusive spreading. (c) Density profile 
in the right outer regime (\emph{i.e.}, sites that are not connected to dephasing probes) with the $x$-axis scaled as $(i\!-\!i_0)/\sqrt{t}$. The perfect collapse of the front (indicated by the downward arrow) indicates the diffusive spreading of the density front in the outer regime. We take $\lambda \!=\! 1.0$. We choose the lattice size $N \!=\! 601$, with $121$ filled lattice sites which are subjected to $N_p\!=\!121$ dephasing probes. The blue dashed line represents the initial density profile for the lattice.}
\label{cric-fig}
\end{figure}

\begin{figure}
\centering
\includegraphics[scale=0.5]{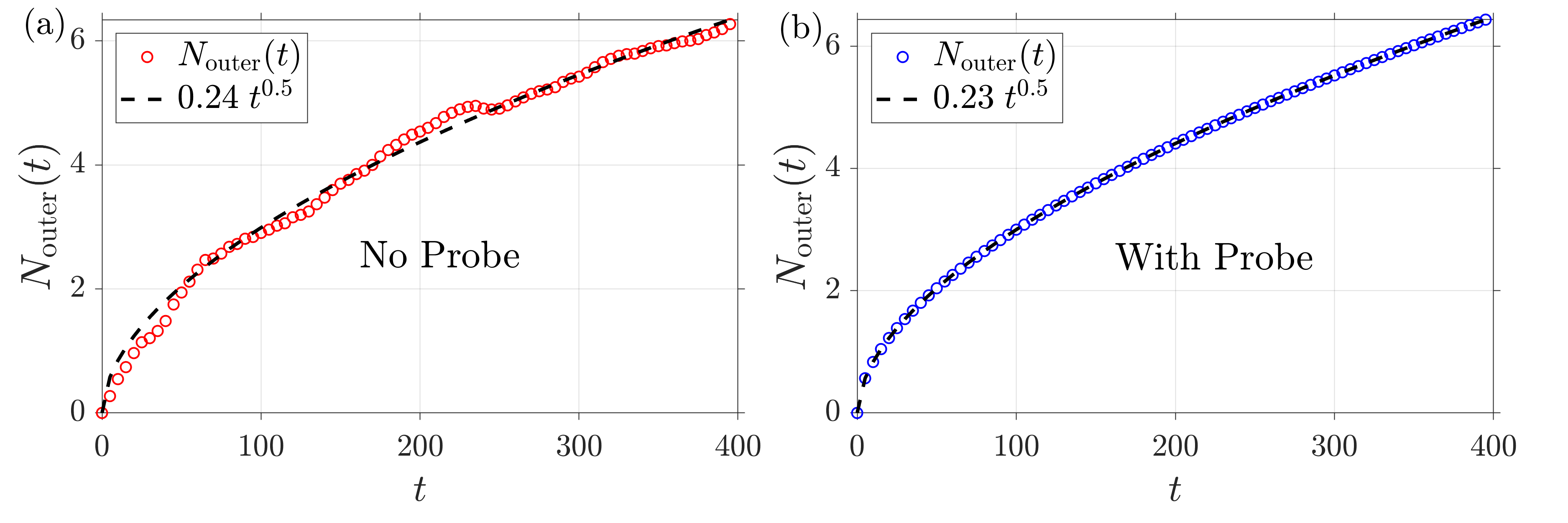}
\caption{$\bm{[}$\textbf{Critical regime}$\bm{]}$ (a) Plot for the total number of electron $N_{\rm outer}(t)$, given in Eq.~\eqref{outer}, as a function of $t$ in the outer regime in the absence of probes. (b)  Plot for the total number of electron $N_{\rm outer}(t)$, given in Eq.~\eqref{outer}, as a function of $t$ in the outer regime in the presence of dephasing probes.  It is clearly seen that $N_{\rm outer}(t)$ has a power-law behavior in time in both in presence and absence of the dephasing probes, \emph{i.e.}, $N_{\rm outer}(t) \!\propto\! t^b$. We find exponent $b \!\sim\! 0.5$ with and without probes. However, in the absence of probes, the profile shows oscillatory behavior which is rooted in the presence of the coherences that are washed away when the probes are present (see \ref{appendix_A} for the details).}
\label{cric-fig-outer}
\end{figure}
We next analyze the inner and the outer regime density fronts as shown in Figs.~\ref{de-loc-fig}(b) and \ref{de-loc-fig}(c), respectively. In Fig.~\ref{de-loc-fig}(b) a perfect collapse of the depletion front is observed upon scaling the $x$-axis as $(i\!-\!i_0)/\sqrt{t}$ where $i_0$ is the location of the right edge of the initial step. This is a clear indication of a diffusive phase. Likewise, in Fig.~\ref{de-loc-fig}(c) a perfect collapse of the growing front is observed upon scaling the $x$-axis as $(i\!-\!i_0)/t$. This is a clear indication of a ballistic phase. Therefore, the setup hosts a mixed diffusive-ballistic dynamical phase. It is important to emphasize that such mixed dynamical phases are the results of the intrinsic interplay between the nature of the single-particle states of the underlying lattice and the suitably placed thermodynamically large number of dephasing probes. 

In Fig.~\ref{de-loc-fig-outer} we have demonstrated the quantum dynamics for the total number of electron $N_{\rm outer}(t)$, given in Eq.~\eqref{outer}, as a function of $t$ in the outer regime (recall schematic Fig.~\ref{Schematic}). Before discussing the open quantum setup addressed in this work, we compute the $N_{\rm outer}(t)$ for the isolated lattice (\emph{i.e.}, in the absence of probes). For this case, one would expect the front to propagate ballistically. Furthermore, in the isolated case, the inner front should also exhibit ballistic propagation. Given this fact and the fact that the areas in the outer and in the inner regimes are the same, leads us to conclude that $N_{\rm outer}$ should be proportional to $t$ which has been verified in Fig.~\ref{de-loc-fig-outer}(a).

Interestingly, when there is a thermodynamically large number of dephasing probes present, a markedly different behaviour $N_{\rm outer}(t)$ is observed. Before proceeding further, we provide a heuristic calculation for what is to be expected. Let $h(t)$ be the typical height of the profile in the outer regime at a given time $t$. Since the spread of the profile in the outer regime is proportional to $t$, the typical area covered in the outer regime is given by $A_{\rm outer}\! \sim\! h(t) \,t$. In a similar way, in the inner regime, the typical magnitude of depletion is $1-h(t)$ and the spread of the depletion front is $\sqrt{t}$. Therefore, the area of the depleted inner region is given by $A_{\rm inner} \!\sim\! \big(1\!-\! h(t)\big)\, \sqrt{t}$. Given that these two areas need to be the same ($A_{\rm outer} = A_{\rm inner} $), we can estimate the typical height $h(t) \!\sim\! \frac{1}{1+\sqrt{t}}$. Therefore, the typical area of the outer regime is $N_{\rm outer} \!\sim\! \frac{t}{1+\sqrt{t}}$. These heuristic estimations for the typical height and the area are consistent with our numerical findings. At very long times, one would expect $N_{\rm outer} \!\sim\! \frac{t}{1+\sqrt{t}} \!\sim\! \sqrt{t}.$ Confirmation of this long-time behavior is depicted in the inset of Fig.~\ref{de-loc-fig-outer}(b). However at intermediate times [see Fig.~\ref{de-loc-fig-outer}(b) ], a power-law of type $N_{\rm outer}(t) \!\propto\!\, t^{b}$ with $b \approx 0.6$ seems to be a reasonable fit and in the very long time limit, we expect $b\to 0.5$.


\subsection{Critical phase \mathinhead{\lambda/J \!=\!1}{}:}
\label{sub-cric}
In this sub-section, we demonstrate the role of dephasing probes on quantum dynamics when the underlying AAH lattice is at the critical point, \emph{i.e.}, $\lambda/J \!=\!1$. In Fig.~\ref{cric-fig}(a) we plot the local density profile $n_i(t)$ and observe a diffusive spreading of the front both in the outer and the inner regime. This diffusive spreading is clearly presented in Figs.~\ref{cric-fig}(b) and \ref{cric-fig}(c). We observe a perfect collapse of the fronts in both regimes by scaling the $x$-axis to $(i\!-\!i_0)/\sqrt{t}$. It is important to emphasize that while both the inner and the outer regimes exhibit diffusive spreading, the diffusion constant is expected to be different. This is because the inner regime experiences diffusion both due to the critical nature of the isolated system and the dephasing probes. On the other hand, the outer regime shows diffusive behavior solely emerging from the critical nature of the AAH lattice. Interestingly, much like the delocalized case, we observe oscillations in the outer regime and no oscillations in the inner regime. In Fig.~\ref{cric-fig-outer} we have demonstrated the quantum dynamics for $N_{\rm outer}(t)$. Both with and without probes we observe $N_{\rm outer}(t) \propto \sqrt{t}$.

\subsection{Localized phase \mathinhead{\lambda/J >1}{}:}
\label{localized}
\begin{figure}
\begin{center}
\includegraphics[scale=0.6]{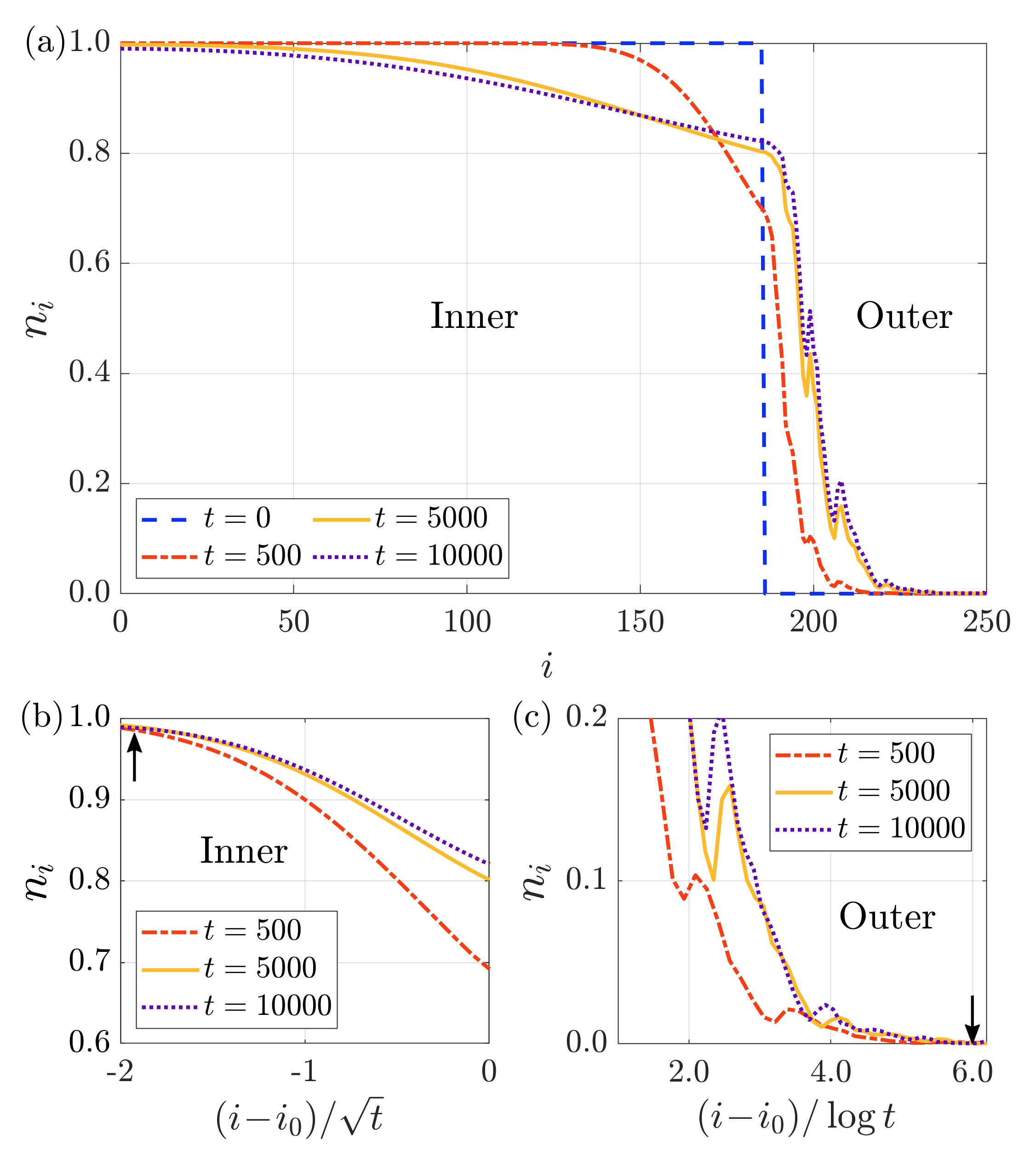}
\end{center}
\caption{$\bm{[}$\textbf{Localized regime}$\bm{]}$ (a) Plot for the local density profile $n_i(t)$, given in Eq.~\eqref{local-density}, as a function of lattice site $i$ for different time snapshots for AAH lattice in the localized phase, $\lambda=1.1$, in the presence of dephasing probes. (b) Density profile in the inner regime (\emph{i.e.}, sites that are connected to dephasing probes) with the $x$-axis scaled by $(i\!-\!i_0)/\sqrt{t}$ to confirm the diffusive spreading of the depletion front. (c) Density profile in the outer regime with the $x$-axis scaled by $(i\!-\!i_0)/\log{t}$. The collapse hints at the logarithmic spreading of the growing front. We choose the lattice size $N \!=\! 501$, with $361$ filled lattice sites which are subjected to $N_p\!=\!361$ dephasing probes. The blue dashed line represents the initial density profile for the lattice. Note that we choose a reasonably large number of initially filled sites in this case, due to the fact that within the time scales of consideration, the density front in the inner regime does not reach the boundary.}
\label{localized-plot}
\end{figure}

\begin{figure}
\centering
\includegraphics[scale=0.7]{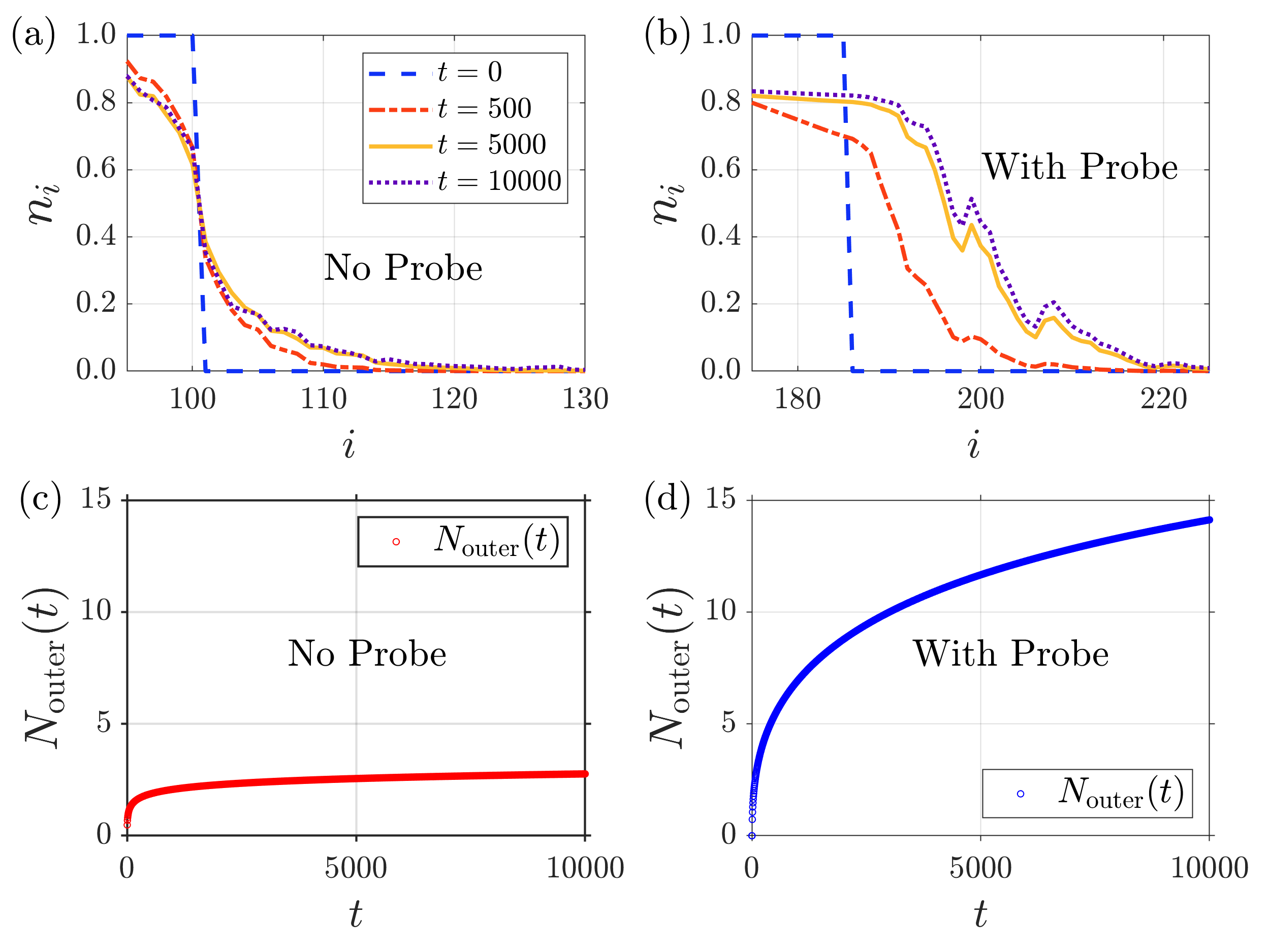}
\caption{$\bm{[}$\textbf{Localized regime}$\bm{]}$ 
(a) Plot for the density profile $n_i(t)$ as a function of lattice sites $i$ for different time snapshots for AAH lattice in the localized phase, $\lambda =1.1$, in the absence of dephasing probes. (b) Plot for density profile $n_i(t)$ as a function of lattice site $i$ for different time snapshots for AAH lattice in the localized phase in the presence of probes. In this case, we note the significant filling up of electrons at the same time snapshot in comparison to (a), where there are no probes. (c) Plot for the total number of electrons $N_{\rm outer}(t)$, given in Eq.~\eqref{outer}, as a function of $t$ in the outer regime in the absence of the dephasing probes. (d) Plot for the total number of electrons $N_{\rm outer}(t)$, given in Eq.~\eqref{outer}, as a function of $t$ in the outer regime in the presence of the dephasing probes. It is observed that $N_{\rm outer}(t)$ gets significantly suppressed when no probes are present. In contrast, in the presence of the probes, diffusive dynamics dominates the inner regime, facilitating faster depletion dynamics. Therefore, accordant with the total electron number conservation,  an enhanced $N_{\rm outer}(t)$ is  reported in (d).}
\label{loc-fig-outer}
\end{figure}

In this subsection, we discuss the impact of dephasing probes when the AAH lattice is in the localized phase, \emph{i.e.}, $\lambda/J \!>\!1$. In Fig.~\ref{localized-plot}(a) we plot the local density profile $n_i(t)$, given in Eq.~\eqref{local-density}, and observe a diffusive spreading in the inner regime, whereas, in the outer regime, interestingly, the spread of the front hints at a logarithmic behaviour in time. 
 We further plot in Figs.~\ref{localized-plot}(b) and \ref{localized-plot}(c), the collapse of these inner and outer fronts by performing appropriate scaling of the $x$-axes. Therefore Fig.~\ref{localized-plot} clearly shows a diffusive-logarithmic mixed phase. The diffusive phase is characterized by $\sqrt{t}$ spreading. Our data indicates that the logarithmic-like phase is characterized by $\log{t}$ spreading (which is therefore extremely slow and almost localized). 
In Fig.~\ref{loc-fig-outer}, we compare the density profiles for the same time snapshots between the cases when dephasing probes are absent [see Fig.~\ref{loc-fig-outer}(a)] and present [see Fig.~\ref{loc-fig-outer}(b)].  We notice a more significant filling up of electrons in the outer regime at the same time snapshot in the presence of probes, in comparison to that in the absence of probes. This filling up of electrons in Fig.~\ref{loc-fig-outer}(b) is caused by the thermodynamically large number of dephasing probes present in the inner regime. The spread of the density profile in the presence of probes is logarithmic in nature whereas one in the absence of probes, the spread is finite with a value given by localization length $\xi \!=\! 1/\ln(\lambda)$ \cite{aubry1980analyticity, PGHarper_1955}.  We next compute the quantum dynamics for $N_{\rm outer}(t)$, given in Eq.~\eqref{outer}. We observe a significant difference in the time dynamics of $N_{\rm outer}(t)$ in the absence [see Fig.~\ref{loc-fig-outer}(c)] and in the presence [see Fig.~\ref{loc-fig-outer}(d)] of probes. In particular, we note a substantial suppression in $N_\mathrm{outer}(t)$ when no probes are introduced. Conversely, with the inclusion of probes, diffusive dynamics take precedence in the inner regime, accelerating the depletion process. As a result, in accordance with the conservation of total electron count, an enhanced $N_\mathrm{outer}(t)$ is evident in Fig.~\ref{loc-fig-outer}(d). 


\section{Summary}
\label{sum}
In summary, we study the impact of dephasing probes on the quantum dynamics of density profiles across a one-dimensional incommensurate AAH fermionic lattice. In the presence of a thermodynamically large number of probes, we report the coexistence of mixed phases (\emph{i.e.}, diffusive-ballistic, diffusive-diffusive, diffusive-localized), due to the remarkable nature of the single-particle eigenstates of the underlying AAH lattice. We characterize these phases by analyzing the spread of the local density front $n_i(t)$, as shown in Fig.~\ref{de-loc-fig} (delocalized), Fig.~\ref{cric-fig} (critical), and Fig.~\ref{localized-plot} (localized). An interesting observation we noted is that the outer regime in Fig.~\ref{de-loc-fig}, Fig.~\ref{cric-fig}, and Fig.~\ref{localized-plot}, exhibits persistent oscillatory features, which remains even after phase averaging,  while the inner regime appears smooth. The reason for this turns out to be rather subtle. Additional investigations (in \ref{appendix_A}) suggest that  the two-point correlation matrix $C_{ij}$, defined in Eq.~\eqref{eq:coherence}, displays a diagonal pattern where the probes are present (\emph{i.e.}, in the inner regime), reflecting the fingerprint of destruction of coherences caused by the dephasing probes. This dephasing effect suppresses the unitary oscillations in the inner regime, and explains the smooth density profile observed in this regime. On contrary to this, in the outer regime, where no probes are attached (and hence, no dephasing effect), coherences develop and persist. As a result, the outer regime shows oscillations. The same line of reasoning applies to both critical and localized regimes. 
We further analyze the growth in the number of electrons $N_{\rm outer}$ within the outer regime across all three phases [see Fig.~\ref{de-loc-fig-outer} (delocalized),  Fig.~\ref{cric-fig-outer} (critical), and Fig.~\ref{loc-fig-outer} (localized)] and showcase significant disparities in our observations when thermodynamically large number of dephasing probes are introduced. 

Although we use the AAH model as our underlying platform, our findings are easily extendable to other kinds of incommensurate models, including those that host mobility edges~\cite{biddle2010predicted, sarma1988mobility, ganeshan2015nearest,li2015many, li2020mobility, deng2017many, purkayastha2017nonequilibrium, roy2021population}, which have also gained importance from an experimental perspective recently~\cite{luschen2018single,kohlert2019observation, an2021interactions}. Our work is an important step forward in terms of engineering physical systems {with various geometries that could host mixed dynamical phases.}

\ack
\label{ackw}
S.M. acknowledges financial support from the CSIR, India (File number: 09/936(0273)/2019-EMR-I). M.K. would like to acknowledge support from Project 6004-1 of the Indo-French Centre for the Promotion of  Advanced Research (IFCPAR), SERB Matrics Grant (MTR/2019/001101) and VAJRA faculty scheme (No.~VJR/2019/000079) from the Science and Engineering Research Board (SERB), Department of Science and Technology, Government of India. M.K. acknowledges support from the Department of Atomic Energy, Government of India, under Project No. RTI4001. 
M.K. thanks the hospitality of the Department of Mathematics of the Technical University of Munich, Garching (Germany). B.K.A acknowledges the MATRICS grant MTR/2020/000472 from SERB, Government of India. B.K.A also thanks the Shastri Indo-Canadian Institute for providing financial support for this research work in the form of a Shastri Institutional Collaborative Research Grant (SICRG). B.K.A. would also like to acknowledge  funding  from  National  Mission  on  Interdisciplinary Cyber-Physical  Systems  (NM-ICPS)  of  the  Department  of Science and Technology, Government Of India, through the I-HUB Quantum Technology Foundation, Pune, India. The authors would like to thank the International Centre for Theoretical Sciences (ICTS) for organizing the program - Periodically and quasi-periodically driven complex systems (code: ICTS/pdcs2023/6) where many interesting discussions related to this project took place.

\appendix 
\section{Coherences and population in systems subjected to probes}
\label{appendix_A}
\begin{figure}
\begin{center}
\includegraphics[width = 1.0\columnwidth]{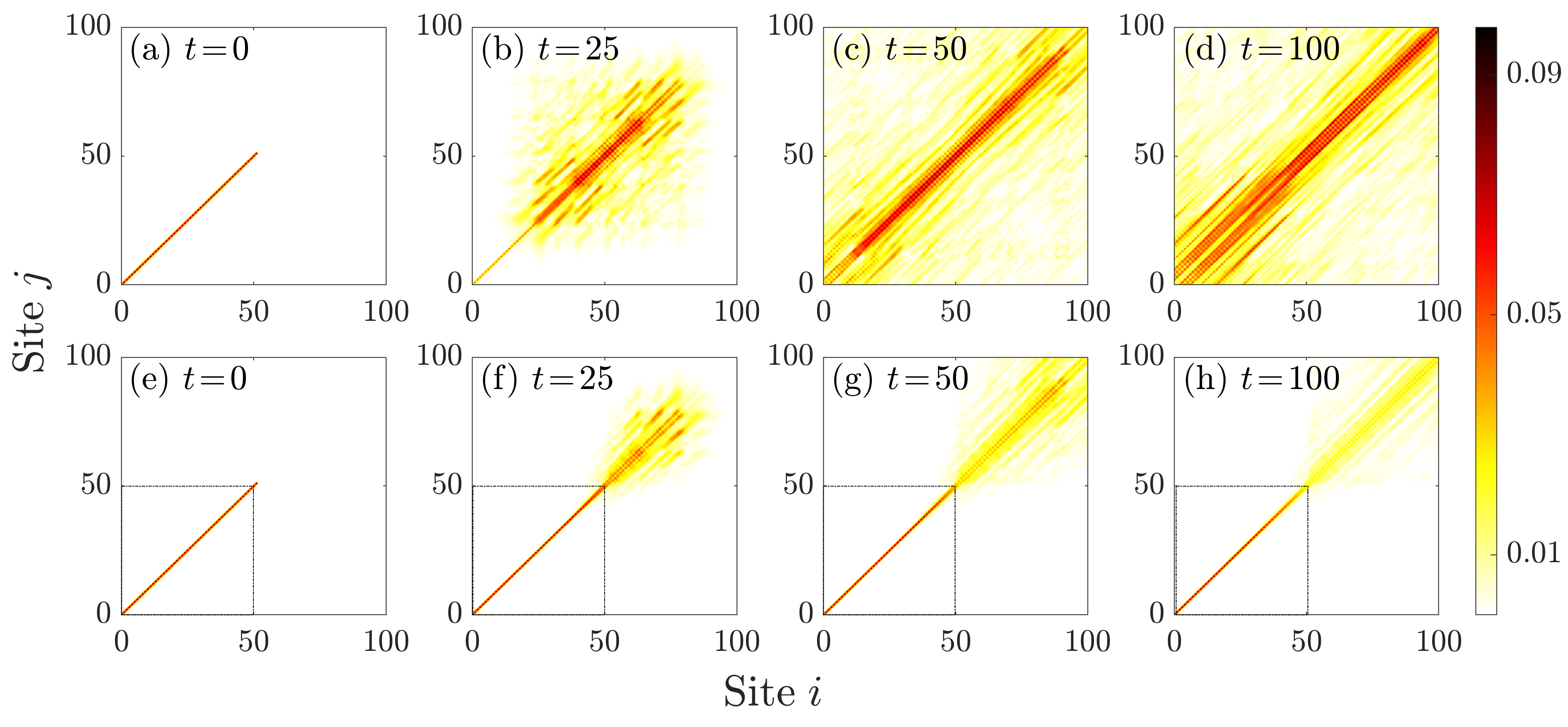}
\caption{$\bm{[}$\textbf{No probe case (a)-(d)}$\bm{]}$  Plot showing the spread of coherences and population $\langle c_i^{\dagger} c_j\rangle$ for different time snapshots (given in the legends) in the delocalized phase, $\lambda=0.5$, of the AAH lattice when no probes are attached. $\bm{[}$\textbf{With probe case (e)-(h)}$\bm{]}$ Plot showing the spread of  coherences and population when the lattice is subjected to a thermodynamically large number of dephasing probes (see schematic Fig.~\ref{Schematic}). The values of the populations i.e., the diagonal elements in all figures are rescaled by a factor of $0.1$ for clarity of presentation. The square boxes in the lower panel (e)-(h) indicate the inner regime that contains the dephasing probes. Therefore within the square box the structure is diagonal.}
\label{app}
\end{center}
\end{figure}

In this appendix, we analyze the coherences in the AAH lattice setup given in schematic Fig.~\ref{Schematic}. We recall that the coherences are defined as $\langle c_i^{\dagger} c_j \rangle$ with $i \!\neq\! j$ and the populations are defined as $\langle c_i^{\dagger} c_i\rangle$. In Figs.~\ref{app}(a)-\ref{app}(d) we show the spread of coherences when no probes are attached to the lattice for four different time snapshots (including the initial time). We observe the development of coherences as the density profile spreads within the lattice. However, in the presence of probes, in Fig.~\ref{app}(e)-\ref{app}(h) coherences are destroyed in the regions where the dephasing probes are attached, \emph{i.e.}, $i\!\leq\!50$, and $ j\!\leq\!50$. In other words, Figs.~\ref{app}(f)-\ref{app}(h) are ``partially diagonal''. On the contrary, in the regimes devoid of probes, \emph{i.e.}, $i\!>\!50$, and $ j\!>\!50$, coherences do persist. Therefore, the lower panel of Fig.~\ref{app} clearly demonstrates the existence of the mixed phase. In addition, the disruption of coherences in presence of dephasing probes results in smooth density profiles in the inner regime, as seen in Fig.~\ref{de-loc-fig}, whereas, the outer regime displayed oscillations as populations are typically coupled to coherences, [for example, see Eq.~\eqref{corr}]. Although our analysis is in the delocalized regime, this analysis straightforwardly applies to both the critical and the localized regimes. 

\newpage
\section*{References}
\bibliographystyle{ieeetr}
\bibliography{references.bib}

\end{document}